\documentstyle[aps,prl,epsf]{revtex}
\def\lsim{\raise0.3ex\hbox{$<$\kern-0.75em\raise-1.1ex\hbox{$\sim$}}}
\def\gsim{\raise0.3ex\hbox{$>$\kern-0.75em\raise-1.1ex\hbox{$\sim$}}}

\newcommand{\beqn} {\begin{equation}}
\newcommand{\eqn} {\end{equation}}
\def\o{\over}

\def\PR{{ Phys.\ Rev.\ }}

\def\NP{{ Nucl.\ Phys.\ }}

\begin{document}

\draft
\tightenlines

\title{
%
%
\begin{flushright}
\normalsize
UTHEP-402 \\
UTCCP-P-64 \\
\end{flushright}
%
%
        Equation of state for pure SU(3) gauge theory \\
        with renormalization group improved action}

\author{CP-PACS Collaboration \\[1mm]
$^1$M.~Okamoto, $^2$A.~Ali Khan, $^1$S.~Aoki, 
$^{1,2}$R.~Burkhalter, $^2$S.~Ejiri,
$^3$M.~Fukugita, $^4$S.~Hashimoto, $^{1,2}$N.~Ishizuka, $^{1,2}$Y.~Iwasaki, 
$^{1,2}$K.~Kanaya, $^2$T.~Kaneko, 
$^5$Y.~Kuramashi%
\thanks{On leave from Institute of Particle and Nuclear Studies,
High Energy Accelerator Research Organization(KEK),
Tsukuba, Ibaraki 305-0801, Japan},
$^2$T.~Manke, $^2$K.~Nagai,
$^4$M.~Okawa, $^{1,2}$A.~Ukawa, $^{1,2}$T.~Yoshi\'e \\[2mm]}

\address{
$^1$Institute of Physics, University of Tsukuba, \\
Tsukuba, Ibaraki 305-8571, Japan \\
$^2$Center for Computational Physics, University of Tsukuba, \\
Tsukuba, Ibaraki 305-8577, Japan \\
$^3$Institute for Cosmic Ray Research, University of Tokyo, \\
Tanashi, Tokyo 188-8502, Japan \\
$^4$High Energy Accelerator Research Organization (KEK), \\
Tsukuba, Ibaraki 305-0801, Japan \\
$^5$Department of Physics, Washington University, \\
St. Louis, Missouri 63130, USA
}

\date{\today}

\maketitle

\begin{abstract}

A lattice study of the equation of state for pure SU(3) gauge theory
using a renormalization-group (RG) improved action is presented.
The energy density and pressure are calculated  
on a $16^3\times 4$ and a $32^3\times 8$ lattice employing the integral 
method.  Extrapolating the results to the continuum limit, we find the 
energy density and pressure to be in good agreement with those 
obtained with the standard plaquette action within the 
error of 3--4\%. 

\end{abstract}

\newpage

\section{Introduction}
\label{sec:intro}

At sufficiently high temperatures, 
the quark-confinement property of QCD is expected to be lost so that 
hadrons dissolve into a plasma of quarks and gluons.
This quark-gluon plasma state must have existed in the early Universe, 
and its experimental detection is  being  actively pursued
through relativistic heavy-ion collisions.  
Basic information that is required 
to explore physical phenomena in the quark-gluon 
plasma is its equation of state, namely the energy density and pressure 
as a function of temperature.  For this reason a number of lattice QCD studies 
of equation of state have been made\cite{reviews}.  
An important progress in this effort is the recent work of the Bielefeld 
group\cite{EOSbielefeld} in which a systematic continuum extrapolation 
was carried out for the equation of state for pure SU(3) gauge theory. 
Using the standard plaquette action,  
they calculated bulk thermodynamic quantities on lattices with the temporal 
extent $N_t = 4$, 6 and 8, 
and extrapolated the results to the continuum limit $N_t \rightarrow \infty$,
assuming that the data at $N_t = 6$ and 8 follow the leading extrapolation 
formula.

To understand the quark gluon plasma in the real world, 
this work has to be extended to full QCD 
with dynamical quarks.  This is a difficult task due to significant 
increase of the amount of computations needed for full QCD simulations. 
One approach to lessen the computational cost is to employ improved 
actions designed to have reduced lattice cut-off effects, and hence should 
allow reliable continuum extrapolation from coarser lattice spacings compared 
to the case for the standard unimproved actions. 
Thermal properties of several improved pure 
gauge actions have already been 
studied\cite{ourPot97,bielefeld22,bielefeld12,papa}.
At present, however, no extrapolation of thermodynamic quantities 
to the continuum limit has been made with improved actions.

In this article we report on our study of the continuum limit of equation 
of state for pure SU(3) gauge theory with an improved action determined from
an approximate renormalization-group argument\cite{Iwasaki83}.
Simulations are carried out on $16^3\times 4$ and $32^3\times 8$ lattices, 
and the energy density and pressure are calculated by the integral 
method\cite{Eng90}. 
We find that the results extrapolated to the continuum limit agree well with 
those obtained from the standard action in Ref.~\cite{EOSbielefeld}.
This provides us with a cross-check of the final results 
in the continuum limit,
and also provide support for the validity of assumptions behind 
the extrapolation procedures.

This paper is organized as follows.
In Section \ref{sec:method},
we summarize the basic formulation and our notations. Some details of 
our simulations are given in Section \ref{sec:simulation}. 
We define our choice of the temperature scale in terms of 
the string tension in Section \ref{sec:Tscale}, 
and examine scaling of the critical temperature
in Section \ref{sec:crit-temp}. 
In Section \ref{sec:eos} we present our results for equation of state 
obtained at $N_t = 4$ and 8, and their continuum extrapolation.
A comparison of our results with those obtained from the standard action
is also made.  In Section \ref{sec:pert} we briefly discuss results 
obtained with the operator method\cite{Engels82}.
We end with a brief conclusion in Section \ref{sec:conclusion}.

\section{Method}
\label{sec:method}

The partition function of a finite temperature SU(3) lattice gauge theory
is defined by 
\beqn
Z  = \int [{\rm d}U]\, e^{~\beta S_g},
\label{eq:FTZ}
\eqn
where $\beta = 6/g^2$ is the bare gauge coupling, and $S_g$
the lattice gauge action.
Denoting the $k\times l$ Wilson loop in the $(\mu, \nu)$-plane at a 
site $x$ as
$
W^{k\times l}_{\mu \nu} (x) = {1 \over 3} {\rm ReTr} 
\biggl[ \prod_{k\times l\,\, {\rm loop}} U \biggr] 
$, 
the renormalization group (RG) improved gauge action we use is 
given by\cite{Iwasaki83}
\beqn
S_g = \;\;c_0\sum_{x,\mu <\nu} W_{\mu\nu}^{1\times 1}(x) 
+c_1\sum_{x,\mu ,\nu} W_{\mu\nu}^{1\times 2}(x)
\label{eqn:6LinkAct} 
\eqn
with $c_0 = 1 - 8c_1$ and $c_1 = -0.331$. 
On a lattice with a size $N_s^3\times N_t$ and the lattice spacing $a$, 
the temperature $T$ and physical volume $V$ of the system are 
given respectively by 
\beqn
T = 1 / (N_t a),
\hspace{8mm}
V = (N_s a)^3~.
\label{eq:FTTemp}
\eqn

We calculate the energy density $\epsilon$ and pressure $p$
using the integral method\cite{Eng90}.
For a large homogeneous system, 
the pressure is related to the free energy density $f$ through
\beqn
p = -f = {T \over V} \ln Z~.  
\label{pressure}
\eqn
Using an identity 
${\partial \over \partial \beta} \ln Z = \langle S_g \rangle$, 
we then have
\beqn
\left. {p\over T^4} \right|^{\beta}_{\beta_0} 
= \int^{\beta}_{\beta_0} {\rm d}\beta' \Delta S~,
\label{freelat}
\eqn
where
\beqn
\Delta S \equiv N_t^4 \bigl( \langle S\rangle_T-\langle S\rangle_0\bigr)~,
\label{eq:DeltaS}
\eqn
with $\langle S\rangle_T$ the expectation value of the action density
$S = S_g / N_s^3 N_t$ at temperature $T$. 
The zero-temperature expectation value $\langle S\rangle_0$ is 
introduced to subtract the vacuum contribution,  
which is conventionally computed on 
a symmetric lattice with the same spatial volume $N_s^3\times N_s$.
Once the pressure is known, the energy density can be computed using 
\beqn
{\epsilon - 3p \over T^4} = T{{\rm d}\beta \over {\rm d}T} \Delta S~,
\label{eq:intmeas}
\eqn
where 
\beqn
T{{\rm d}\beta \over {\rm d}T} = -a{{\rm d}\beta \over {\rm d}a} 
\equiv \beta(g) 
\label{betafunc}
\eqn
is the QCD beta-function.

\section{Simulations}
\label{sec:simulation}

The fundamental quantity used in the integral method is the action 
difference $\Delta S$ defined by (\ref{eq:DeltaS}).
In order to calculate $\langle S\rangle_T$ and $\langle S\rangle_0$,
we need to simulate both asymmetric
($N_s^3\times N_t$) and symmetric ($N_s^3\times N_s $) lattices.
The spatial lattice size $N_s$ should be sufficiently large to suppress
finite size effects.
Past finite-size studies\cite{FOU,QCDPAX,ourPot97} suggest that 
the condition $N_s / N_t$ \gsim\ 3 is the minimum requirement.
As with the work of the Bielefeld group for the plaquette 
action\cite{EOSbielefeld}, we choose $N_s / N_t = 4$ and perform 
simulations on $16^3\times 4$  and $32^3\times 8$ lattices
as well as on $16^4$ and $32^4$ lattices for a set of values of $\beta$
around and above the critical point.

Gauge fields are updated by the pseudo-heat-bath
algorithm with five hits, followed by four 
over-relaxation sweeps; 
the combination of these updates is called an iteration.
We always start from the completely ordered configuration,
and perform $20\,000$ to $36\,000$ iterations after thermalization 
on asymmetric lattices,
and about $10\,000$ iterations on symmetric lattices.  We find these 
number of iterations to be sufficient for a statistical accuracy of 2--3\% 
or better for the value of $\Delta S$.
The statistics of our runs are compiled in 
Table~\ref{tab:stat}.

\begin{figure}[tb]
\begin{center}
\leavevmode
    \epsfxsize=9cm 
    \epsfbox{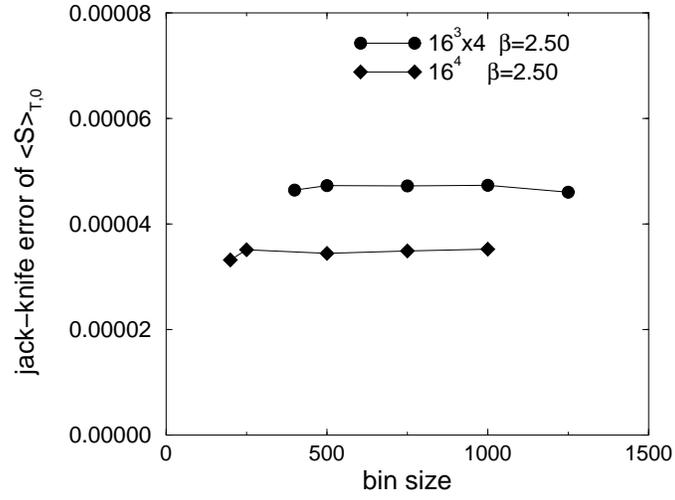}
\end{center}
    \caption{Bin size dependence of the jack-knife error of the action density}
\label{fig:jerror}
\end{figure}

We measure Wilson loops and Polyakov loop at every iteration.
Errors are determined by the jack-knife method.
The typical bin size dependence of the jack-knife error for 
the action density is shown in Fig.~\ref{fig:jerror}. 
The errors are almost constant over a wide range of bin sizes, and 
we adopt the bin size of 1000 iterations for asymmetric lattices
and 500 on symmetric lattices.

In Table~\ref{tab:plaq16}
we list the expectation value of the action density $\langle S\rangle_T$
calculated on asymmetric lattices of size
$16^3\times 4$ and $32^3\times 8$  and that for 
$\langle S\rangle_0$ on symmetric lattices of size $16^4$ and $32^4$.

\section{Temperature scale}
\label{sec:Tscale}

\begin{figure}[t]
\begin{center}
\leavevmode
\epsfxsize=11cm 
    \epsfbox{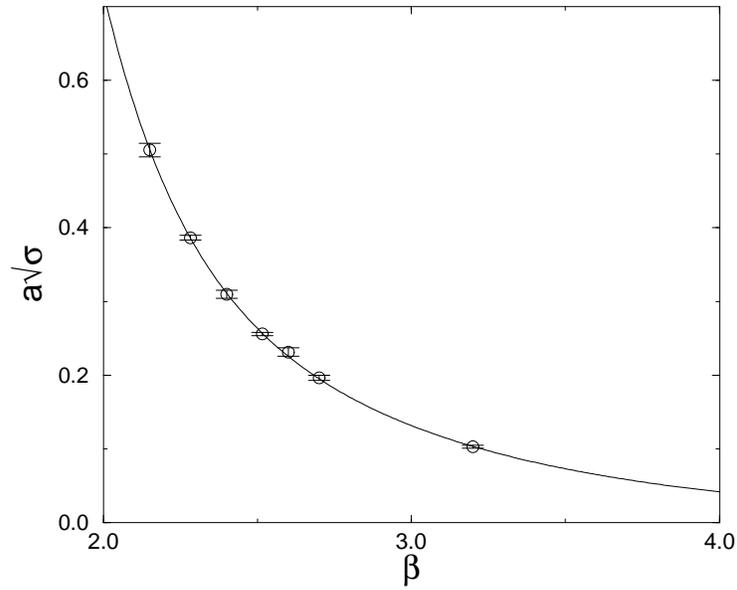}
\end{center}
\caption{String tension as a function of gauge coupling. Solid line 
represents a fit to Allton's parametrization.} 
\label{fig:b-sigma}
\end{figure}

In order to determine the temperature $T = 1/(N_t a(\beta ))$,
we need to compute the lattice 
spacing $a$ as a function of the gauge coupling $\beta$.
We use the string tension of static quark potential to fix the scale of 
this relation.
In Table~\ref{tab:allsig} we compile results for the dimensionless string 
tension $a\sqrt{\sigma}$ obtained with the RG-improved 
action\cite{ourPot97,kaneko}.
We fit these results by an ansatz proposed by Allton\cite{Allton},
\beqn\label{Allton}
(a\sqrt{\sigma})(\beta) =  
                         f(\beta) ~ ( \,
  1 + c_2           \, \hat{a}(\beta)^2
    + c_4           \, \hat{a}(\beta)^4 + \ldots \, 
                            )/ c_0 \, , \qquad
     \hat{a}(\beta) \equiv { f(\beta) \over f(\beta_1)} \, ,
\eqn
where $f(\beta)$ is the two-loop scaling function of SU(3) gauge theory,
\beqn\label{f2loop} 
 f(\beta = 6/g^2) \, \equiv \,
(b_0 g^2)^{-{b_1\o 2b_0^2}} \, \exp(-{1\o 2b_0 g^2}) \, , \qquad
  b_0 = {11\o (4\pi)^2 } \, , \quad  b_1 = {102\o (4\pi)^4} \, ,
\eqn
and $c_n (n=2,4,\cdots)$ parameterize deviations from the two-loop scaling.
Truncating the power corrections at $n=4$ and choosing $\beta_1=2.40$, 
we obtain from this fit,
\beqn
                 c_0  =  0.524(15)\, , \\
                 c_2  =  0.274(76) \, , \\
                 c_4  =  0.105(36) \, 
\eqn
with $\chi^2/dof=0.356$ for 4 degrees of freedom.
As shown in Fig.\ref{fig:b-sigma} the fit curve reproduces the data very 
well.

With this parametrization, the temperature in units of 
the critical temperature $T_c$ is given by 
\beqn
\frac{T}{T_c} = \frac{(a\sqrt{\sigma})(\beta_c)}{(a\sqrt{\sigma})(\beta)}.
\eqn
with $\beta_c$ the critical coupling.  The beta-function is obtained 
by differentiating the left hand side of (\ref{Allton}) with respect 
to the lattice spacing $a$, keeping $\sigma$ constant. 
 
\section{Critical temperature}
\label{sec:crit-temp}

We determine the critical coupling $\beta_c(N_t, N_s)$ 
for the deconfinement transition on an $N_s^3\times N_t$ lattice 
by the peak location of the susceptibility $\chi$ of 
the Z(3)-rotated Polyakov line.
The values of $\beta_c$ for $N_t = 3$, 4, 6 have been reported 
in Ref.~\cite{ourPot97}.
In order to compute $\beta_c$ for $N_t = 8$, 
we perform an additional simulation of $24\,000$ iterations 
at $\beta = 2.710$ on a $32{^3}\times 8$ lattice. 
The $\beta$ dependence of $\chi$ is calculated by 
the spectral density method~\cite{SDM}.
We estimate the error by the jack-knife method with the bin-size of 
2000 iterations. 
The values of $\beta_c(N_t, N_s)$ for finite $N_s$ are summarized 
in the second column of Table~\ref{tab:critical}.

Calculating the critical temperature requires an extrapolation of 
$\beta_c(N_t, N_s)$ toward infinite spatial size $N_s\to\infty$ for 
each $N_t$.  For the first-order transition of the pure gauge system, 
the spatial volume dependence of $\beta_c(N_t, N_s)$ is expected to 
follow\cite{EOSbielefeld}
\beqn
\beta_c(N_t, N_s)=\beta_c(N_t,\infty)-c(N_t)\frac{N_t^3}{N_s^3}.
\label{eq:fss}
\eqn
It has been reported in Ref.~\cite{ourPot97} that results for $N_t=3$ and 4
reasonably satisfy (\ref{eq:fss}) with 
$c(N_t)=0.122(54)$ for $N_t=3$ and $0.133(63)$ for $N_t=4$.  
An approximate scaling of the coefficient $c(N_t)$ motivates us to 
apply (\ref{eq:fss}) for $N_t=6$ (as was made in Ref.~\cite{ourPot97}) 
and also for $N_t=8$, adopting the value  $0.133(63)$ for the coefficient.
Substituting values of $\beta_c(N_t, \infty)$
in the parametrization of the string tension (\ref{Allton}), 
we calculate $T_c/\sqrt{\sigma}=1/(N_ta\sqrt{\sigma})$.  
We tabulate results of this
analysis in the third and fourth column of Table~\ref{tab:critical}.

\begin{figure}[t]
\begin{center}
\leavevmode
    \epsfxsize=11cm 
    \epsfbox{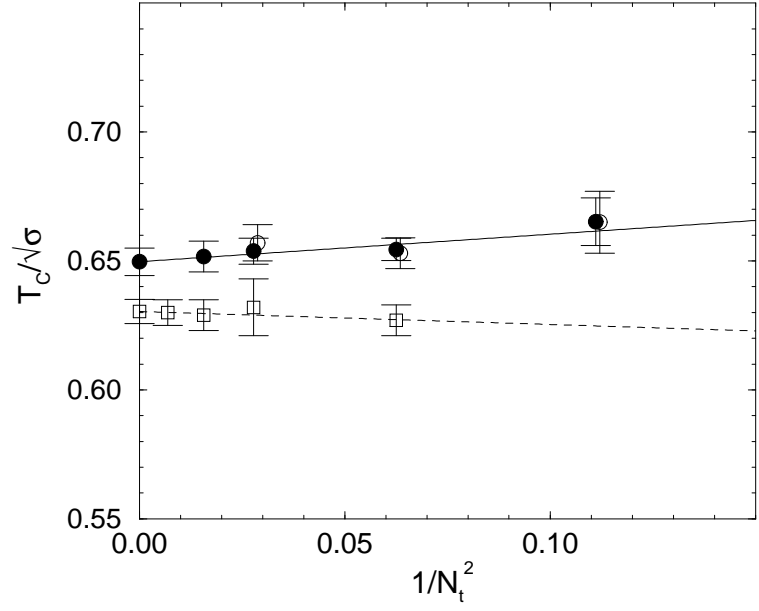}
\end{center}
\caption{
$T_c/\protect\sqrt{\sigma}$ as a function of $1/N_t^2$. Open 
circles are results reported in Ref.~\protect\cite{ourPot97}.  Open 
squares are values for the plaquette action\protect\cite{Beinlich}.} 
\label{fig:Tcscaling}
\end{figure}

In Fig.~\ref{fig:Tcscaling} we plot the results for $T_c/\sqrt{\sigma}$ 
as a function of $1/N_t^2$ (filled circles).  Also shown are the values 
previously reported in Ref.~\cite{ourPot97} (open circles) 
and those for the plaquette action from Ref.~\cite{Beinlich} (open squares).  
A slight difference between the present results and those from 
Ref.~\cite{ourPot97} for the same action 
stems from the fact that an exponential ansatz
$\sqrt{\sigma}a=A\exp (-B\beta)$ with fit parameters $A$ and $B$ 
was adopted in the previous work, which deviates from the parametrization 
(\ref{Allton}).  
We think that the present parametrization 
gives a better estimate of $\sigma$, being 
theoretically consistent with the asymptotic scaling behavior for large 
$\beta$.
The results for the plaquette action is obtained with a 
parametrization\cite{ASDWGA}  
\beqn
\label{sigparam}
(a\sqrt{\sigma})(\beta )  \!=\!  \, f(\beta ) ~ ( \,
  1  + 0.2731      \, \hat{a}(\beta )^2
     - 0.01545     \, \hat{a}(\beta )^4
     + 0.01975     \, \hat{a}(\beta )^6 \, )/0.01364 \, 
\eqn
with $\beta_1=6.0$.

We observe that the new value of $T_c/\sqrt{\sigma}$ for the 
RG-improved action for $N_t=8$ is consistent with the previous results 
for $N_t=3$, 4, 6\cite{ourPot97}.  The difference in this ratio  
obtained for the two actions, however, still remains.  
Making a quadratic extrapolation in $1/N_t$,  
we find $T_c/\sqrt{\sigma}=0.650(5)$ for the RG action, which is 
3\% higher than the value $0.630(5)$ for the plaquette 
action\cite{Beinlich}.  A possible origin of the discrepancy is 
systematic uncertainties in the determination of the string tension
for the two actions, which differ in details.  
We consider that checking an agreement beyond a few percent accuracy, 
as is needed here, would require the generation and analyses of potential 
data over the relevant range of lattice spacings in a completely parallel 
manner for the two actions, which is beyond the scope of the present 
work.

\section{Equation of state}
\label{sec:eos}

\subsection{Results for RG-improved action}
\label{sec:results}

\begin{figure}[t]
\begin{center}
\leavevmode
    \epsfxsize=11cm 
    \epsfbox{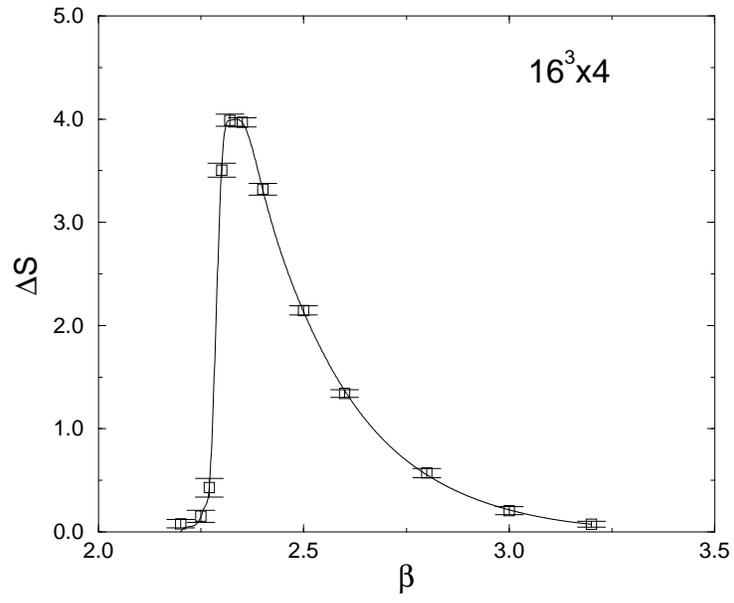}
\vspace{-5mm}
    \epsfxsize=11cm 
    \epsfbox{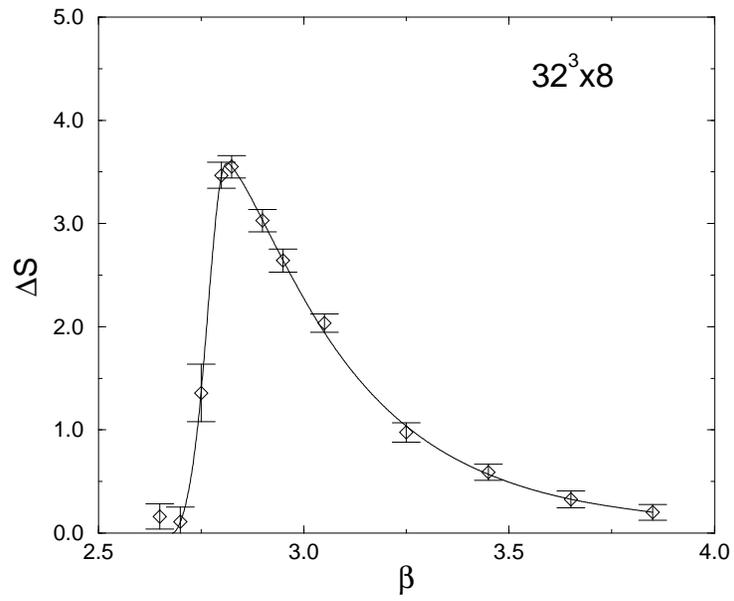}
\end{center}
\caption{Action difference $\Delta S$  for $N_t = 4$ and $8$ as a function 
of $\beta$.}
\label{fig:plaqdif}
\end{figure}

Our results for $\Delta S$ at $N_t = 4$ and 8 are shown in 
Fig.~\ref{fig:plaqdif}. 
In order to integrate $\Delta S$ in terms of $\beta$ to obtain the pressure, 
we have to make an interpolation of the data points.
At large $\beta$ where $T \geq 2T_c$ is satisfied, 
we fit the points by a perturbative ansatz,
\beqn
\Delta S ~=~  {a_2 \o \beta^2} ~+~
{a_3 \o \beta^3} ~+~ {a_4 \o \beta^4} + \cdots,
\eqn
truncating the series at the order $\beta^{-4}$.  The absence of the linear 
term in perturbation theory can be checked easily.
For $\Delta S$ at lower $\beta$-values corresponding to $T \leq 2T_c$, 
we perform a cubic spline fit with the requirement  
that the curve smoothly joins to the large-$\beta$ fit curve at $T = 2T_c$.
The interpolation curves are shown in Fig.\ref{fig:plaqdif}.

\begin{figure}[tb]
\begin{center}
\leavevmode
    \epsfxsize=12cm 
    \epsfbox{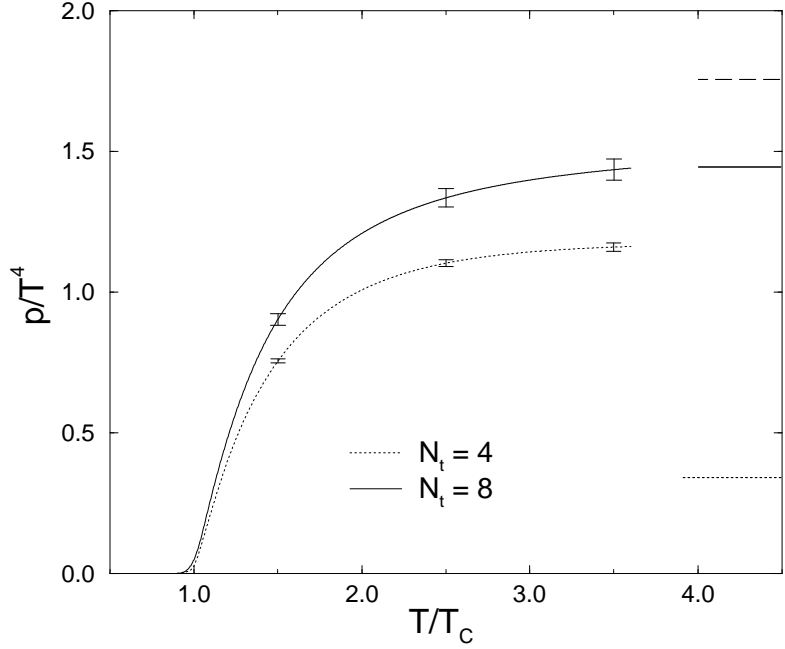}
\end{center}
\caption{Pressure for $N_t = 4$ and 8.
The dashed horizontal line on the top-right 
represents the leading order perturbative value in the high temperature 
limit in the continuum, and 
solid and dotted lines are the corresponding lattice values 
for $N_t =8$ and 4.}
\label{fig:T-P.RG}
\end{figure}

\begin{figure}[b]
\begin{center}
\leavevmode
    \epsfxsize=12cm 
    \epsfbox{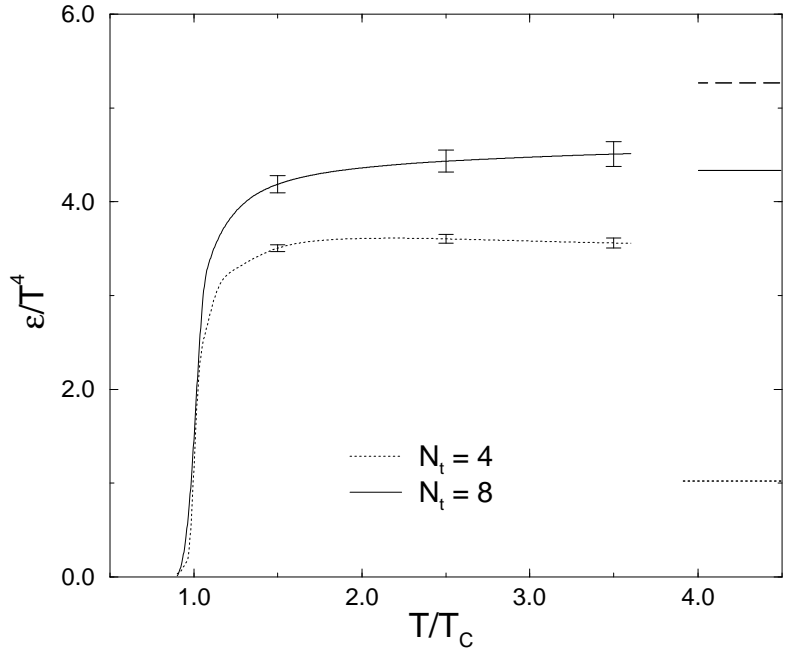}
\end{center}
\caption{Energy density for $N_t = 4$ and 8. 
Meaning of horizontal lines is the same as in
Fig.~\protect\ref{fig:T-P.RG}
}
\label{fig:T-E.RG}
\end{figure}

We use these curves to evaluate the integral for pressure in (\ref{freelat}). 
The lower limit of integration is chosen to be $\beta_0=2.20 ~(N_t=4)$ 
and $2.63 ~(N_t=8)$. 
The results are shown by solid lines in Fig.~\ref{fig:T-P.RG}.
Combining these results for $p$ with those for $\epsilon-3p$ 
computed using (\ref{eq:intmeas}), 
we also obtain the energy density $\epsilon$, which we  
show in Fig.~\ref{fig:T-E.RG}.

The statistical error $\delta p(T)$ for the pressure,  plotted 
at representative points in Fig.~\ref{fig:T-P.RG},  is evaluated 
from the contributions $\delta_i p(T)$ at 
each simulation point $\beta_i$ which is estimated by the jack-knife method.
Since simulations at different $\beta_i$ are statistically independent, 
we compute the final error by the naive error-propagation rule,
$\delta p(T) = \sqrt{\sum_i {\delta p_i(T)^2}}$, 
summing up all the contributions from $\beta_i$ smaller than 
$\beta$ corresponding to the temperature $T$. 
The error $\delta\epsilon (T)$ of the energy density is calculated 
by quadrature from the error of $3\delta p(T)$ and that for 
$\epsilon(T)-3p(T)$,
the latter being proportional to the error of $\Delta S$.

We observe in Figs.~\ref{fig:T-P.RG} and \ref{fig:T-E.RG}
that the energy density and pressure
exhibit a sizable increase between $N_t=4$ and 8.
This increase is opposite to the trend for the plaquette action, but 
it is consistent with the prediction of 
the leading order perturbative result shown by horizontal lines at the 
right of the figure.  The values from the integral method, however, 
overshoot those from perturbation theory toward high temperatures, 
particularly for $N_t=4$ for which the perturbative value is 
quite small. We discuss this point further in 
Sec.~\ref{sec:pert}.

\begin{figure}[tb]
\begin{center}
\leavevmode
    \epsfxsize=13cm 
    \epsfbox{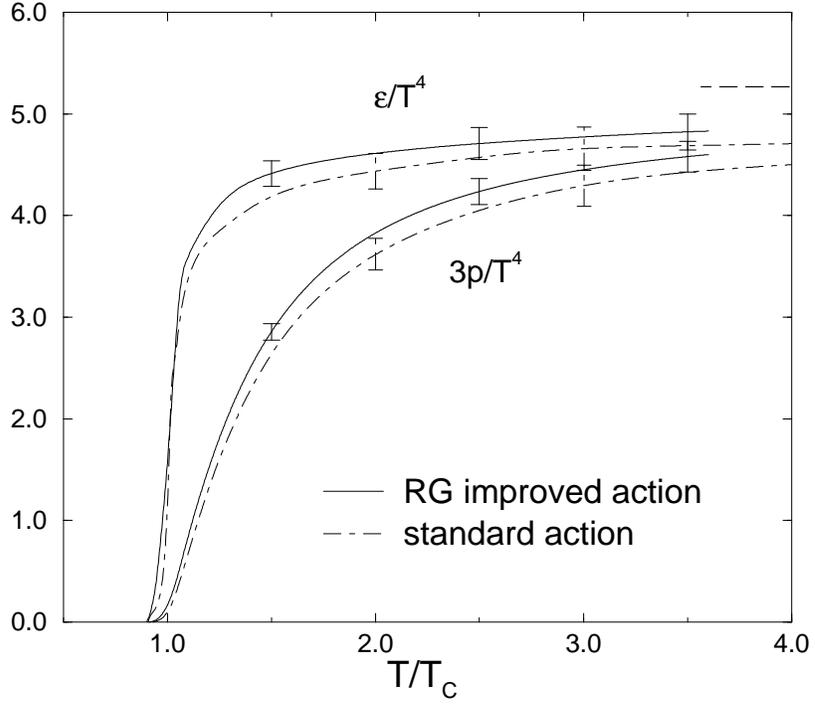}
\end{center}
\caption{Equation of state in the continuum limit for the RG-improved 
action (solid lines) and for the plaquette action 
(dash-dotted lines).  The latter is obtained with the Allton's 
parametrization of string tension using raw data in 
Ref.~\protect\cite{EOSbielefeld}.  
Dashed horizontal line on the top-right 
shows the free gluon gas value in the high temperature limit.}
\label{fig:continuum}
\end{figure}

We now extrapolate the results for energy density and pressure 
to the continuum limit $N_t \rightarrow \infty$.
The RG-improved gauge action has lattice discretization errors of $O(a^2)$. 
Therefore, at a fixed temperature in physical units, 
we expect deviations of thermodynamic quantities 
from the continuum limit to be $O(1/N_t^2)$:
\beqn
\label{eq:contlim}
\biggl({{\cal F} \over T^4}\biggr)_{N_t} 
=~ \biggl({{\cal F} \over T^4}\biggr)_{\rm cont}
+~ {{c_{\cal F}(T) \over N_t^2}}_{~~,}\qquad {\cal F} = p, \epsilon.
\label{cfit}
\eqn
Extrapolating the results for $\epsilon$ and $3p$ at $N_t =4$ and 8 
with this form, we obtain the continuum predictions drawn by solid lines 
in Fig.~\ref{fig:continuum}.

\subsection{Comparison with results for the plaquette action}
\label{sec:comparison}

We compare our results with those of the Bielefeld group 
obtained with the plaquette action\cite{EOSbielefeld}. 
Care is needed in this comparison since they used a scheme different 
from ours to fix the temperature scale: their scheme is based on 
the requirement that the critical temperature $T_c$ is independent 
of the temporal size $N_t$. 

Since a difference in the scale can sizably affect results for 
thermodynamic quantities\cite{Ejiri98}, we first examine 
the possible influence of this difference. 
For this purpose we reanalyze the raw data of Ref.~\cite{EOSbielefeld} for 
the action density employing the scale parametrization (\ref{sigparam}).
The results for $p$ and $\epsilon$ are shown
by solid lines in Fig.~\ref{fig:stresult} for the temporal sizes 
$N_t = 6$ and $8$ used by the Bielefeld group for the continuum 
extrapolation. 
Compared with their original results, drawn by dashed lines, 
the influence of the scale determination 
is well within the statistical error of 1--3\%.

We extrapolate the pressure and energy density obtained with the 
scale (\ref{sigparam})
to the continuum limit according to (\ref{eq:contlim}).
This leads to the dash-dotted curves shown in Fig.~\ref{fig:continuum}. 
Errors are evaluated in the same way as for the case of the RG-improved 
action.
We observe that the curves for the RG-improved action (solid lines) and 
those for the plaquette action (dash-dotted lines) are in good agreement, 
within the error of 3--4\%, over the entire temperature interval shown. 
This is highly non-trivial since results for the two actions differ 
significantly at finite lattice spacings.  

\begin{figure}[t]
\begin{center}
\leavevmode
    \epsfxsize=11cm 
    \epsfbox{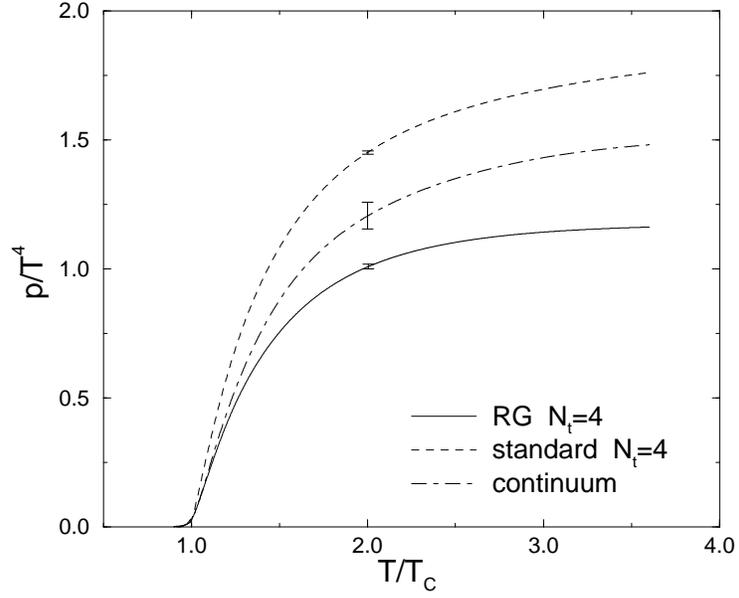}
\end{center}
\caption{Pressure at $N_t = 4$ from the RG and plaquette actions.
For comparison, we also plot the result in the continuum limit
obtained with the plaquette action using 
our choice of the scale discussed in the text.}
\label{fig:T-FT}
\end{figure}

In Fig.~\ref{fig:T-FT} we compare the pressure at $N_t = 4$ 
from the two actions to the result in the continuum limit.
We note that
the continuum results for the RG-improved action are approached from below,
while those for the plaquette action from above.  
We also find that the magnitude of deviation from the continuum
limit is comparable for both actions.
The form of the RG-improved action we employed is determined 
so as to best approximate the renormalized trajectory after a few 
RG transformations, 
within those limited actions with a maximum of 6-link loops. 
Therefore, low-momentum modes, with a momenta smaller than the inverse of 
several lattice spacings, are improved.
On the other hand,
a momentum scale which is significant at high temperatures
is $T = 1/N_t a$ on finite-$N_t$ lattices.
From Fig.~\ref{fig:T-FT},
it appears to be required to add further terms in the action
in order to reduce the cut-off effects for high-momentum modes 
with momentum \gsim\ $1/4a$, 
thereby improving the behavior of the pressure at high temperatures 
on $N_t = 4$ lattices.

\section{Comparison with results with operator method}
\label{sec:pert}

In Sec.~\ref{sec:results} we noted that the 
pressure and energy density calculated 
by the integral method exceed the values corresponding to the perturbative 
high temperature limit.  This is a puzzling result, especially for pressure; 
while $p/T^4$ has to decrease at high temperatures 
to agree with the perturbative result, 
we expect it to be an increasing function of 
temperature since it is given 
by an integral of $\Delta S$ which is generally positive. 
The discrepancy is particularly large for $N_t=4$ for which the leading 
order perturbative results on the lattice are quite small compared to 
those in the continuum as first noted in Ref.~\cite{KarschTsukuba}. 
In Table~\ref{tab:PHL} we list the perturbative value of pressure
on a $N_s^3\times N_t$ lattice in units of the free gluon gas value 
in the continuum for $N_t=4$--12 for the RG-improved and plaquette actions.

In order to further examine this problem, 
we calculate thermodynamic quantities 
in an alternative way using the formulae of the operator 
method\cite{Engels82} given by
\begin{eqnarray}
\frac{\epsilon}{T^4} &=& 
\frac{18}{g^{2}} N_t^4 \bigl[ c_s (g) (\langle S_s \rangle -
\langle S \rangle_0) -c_t (g) (\langle S_t \rangle
-\langle S \rangle_0) \bigr] ,\\
\frac{p}{T^4} &=& \frac{1}{3}\frac{\epsilon}{T^4} - N_t^4
\beta(g) \bigl[\langle S_s + S_t\rangle - 2\langle S
\rangle_0 \bigr] ,
\label{eqn:Pope}
\end{eqnarray}
where $S_s$ and $S_t$ are the spatial and temporal part of the action density,
and the asymmetry coefficients are defined by 
\beqn
c_s (g) = 1 - g^2 \left.{dg_s^{-2} \over d\xi}\right\vert_{\xi=1}, \qquad
c_t (g) = 1 + g^2 \left.{dg_t^{-2} \over d\xi}\right\vert_{\xi=1}.
\label{eqn:KarshCoef}
\eqn
The one-loop values of the asymmetry coefficients for the plaquette 
action have been long known\cite{Karsch}, and preliminary values for 
the RG-improved action have recently been reported\cite{Sakai}.

\begin{figure}[t]
\begin{center}
\leavevmode
    \epsfxsize=11cm 
    \epsfbox{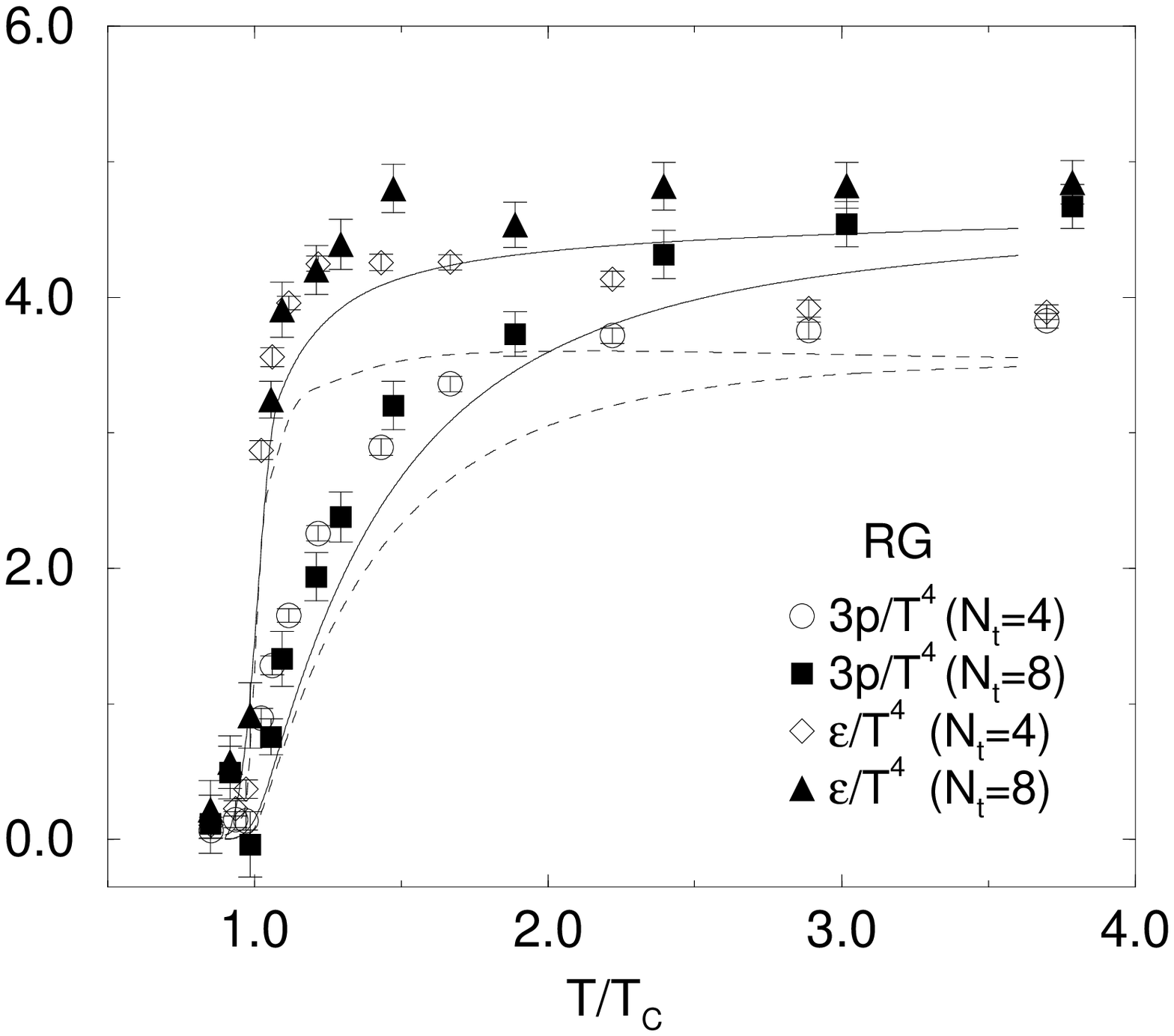}
\vspace{-0.5cm}
    \epsfxsize=11cm 
    \epsfbox{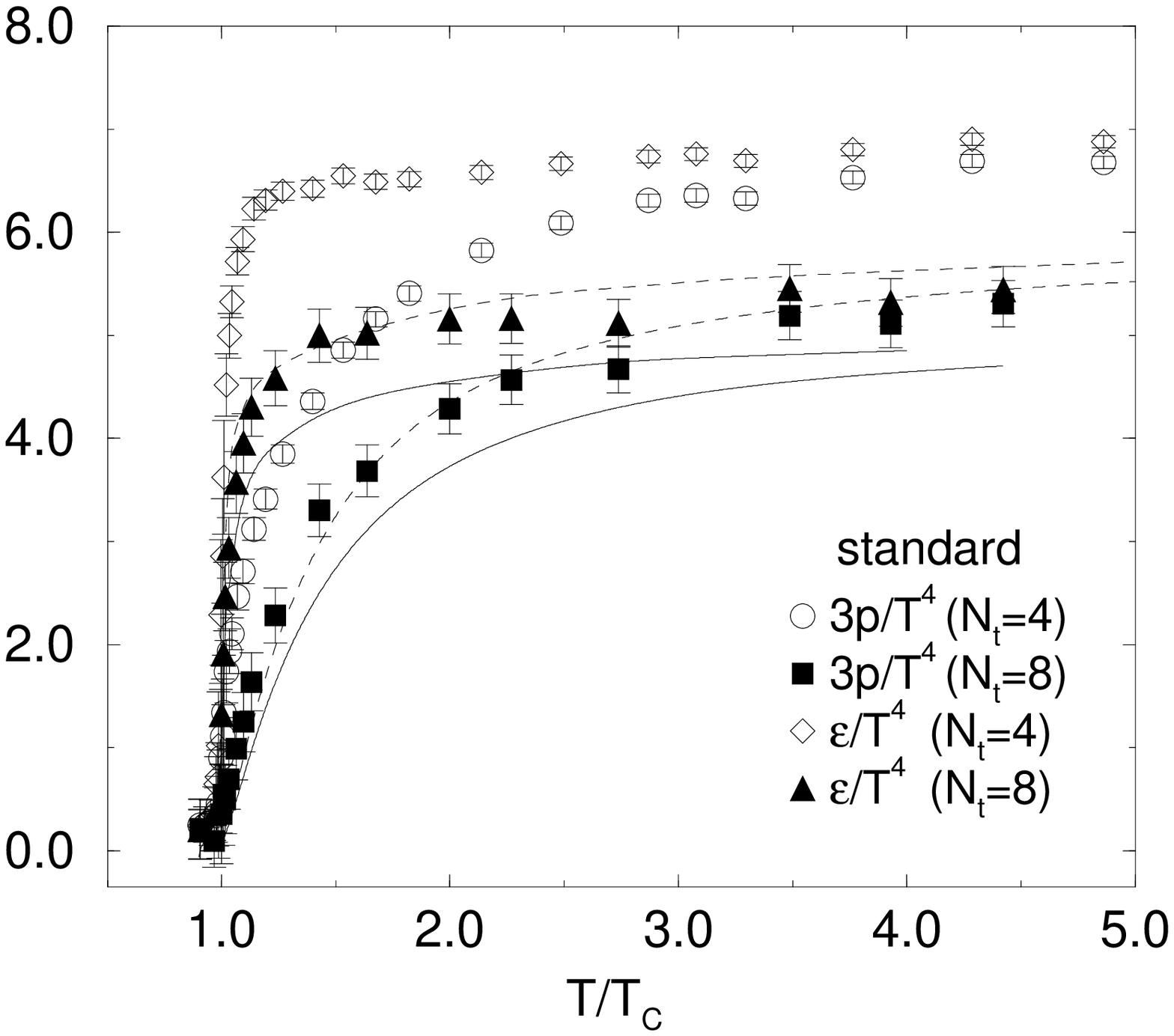}
\end{center}
\caption{Results for energy density and pressure for $N_t=4$ and 8 
obtained with the operator method using one-loop values for 
the asymmetry coefficients
as compared with those from the integral method drawn by
dashed ($N_t=4$) and solid ($N_t=8$) lines. 
Upper figure is for the RG-improved action, and lower figure 
for the plaquette action.}
\label{fig:operator}
\end{figure}

We compare results for the energy density and pressure from 
the integral method and the operator method with one-loop asymmetry 
coefficients in Fig.~\ref{fig:operator} for $N_t=4$ and 8.  
For both types of actions, the values for 
the operator method lie above those for the integral method, and the 
difference diminishes with increasing $N_t$.  
For $N_t=4$ for the RG-improved action, in particular, 
we do not observe any indication of decrease 
toward the perturbative high temperature limit, both with the integral 
and operator methods, at least within the temperature range where we have 
results. 

A possible source of the discrepancy 
is breakdown of perturbation theory 
due to the infrared divergence\cite{Linde} of the theory.  
In the continuum there are 
non-perturbative contributions to free energy beyond 3-loop level.
This problem should also exist on the lattice, and the magnitude of 
non-perturbative contributions may vary depending on the choice of
lattice actions. 

\section{Conclusions}
\label{sec:conclusion}

In this article we have presented results on the equation of state 
for a pure SU(3) gauge theory obtained with an RG-improved gauge action. 
The continuum result for the energy density and pressure show 
an agreement with the results of the Bielefeld group for the plaquette 
action within the error of 3--4\%. 
This provides a concrete support for the expectation that continuum 
results are insensitive to the choice of lattice actions.  

We also found that the energy density and pressure for finite $N_t$ 
overshoot the perturbative high temperature limit. 
Understanding the origin of this behavior shall be explored in the future.

\section*{Acknowledgements}

\begin{figure}[t]
\begin{center}
\leavevmode
    \epsfxsize=10cm 
    \epsfbox{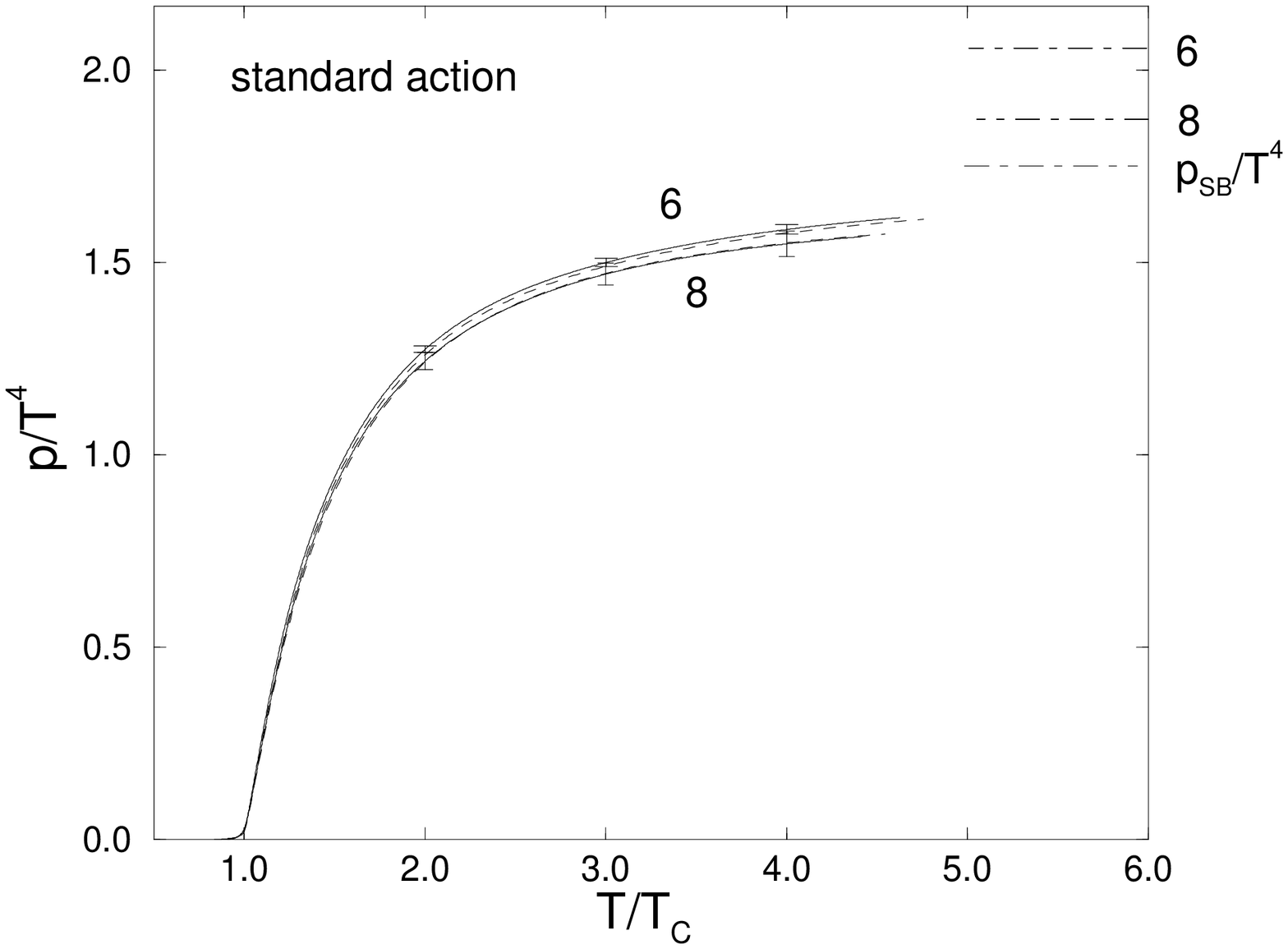}
\vspace{-0.5cm}
    \epsfxsize=10cm 
    \epsfbox{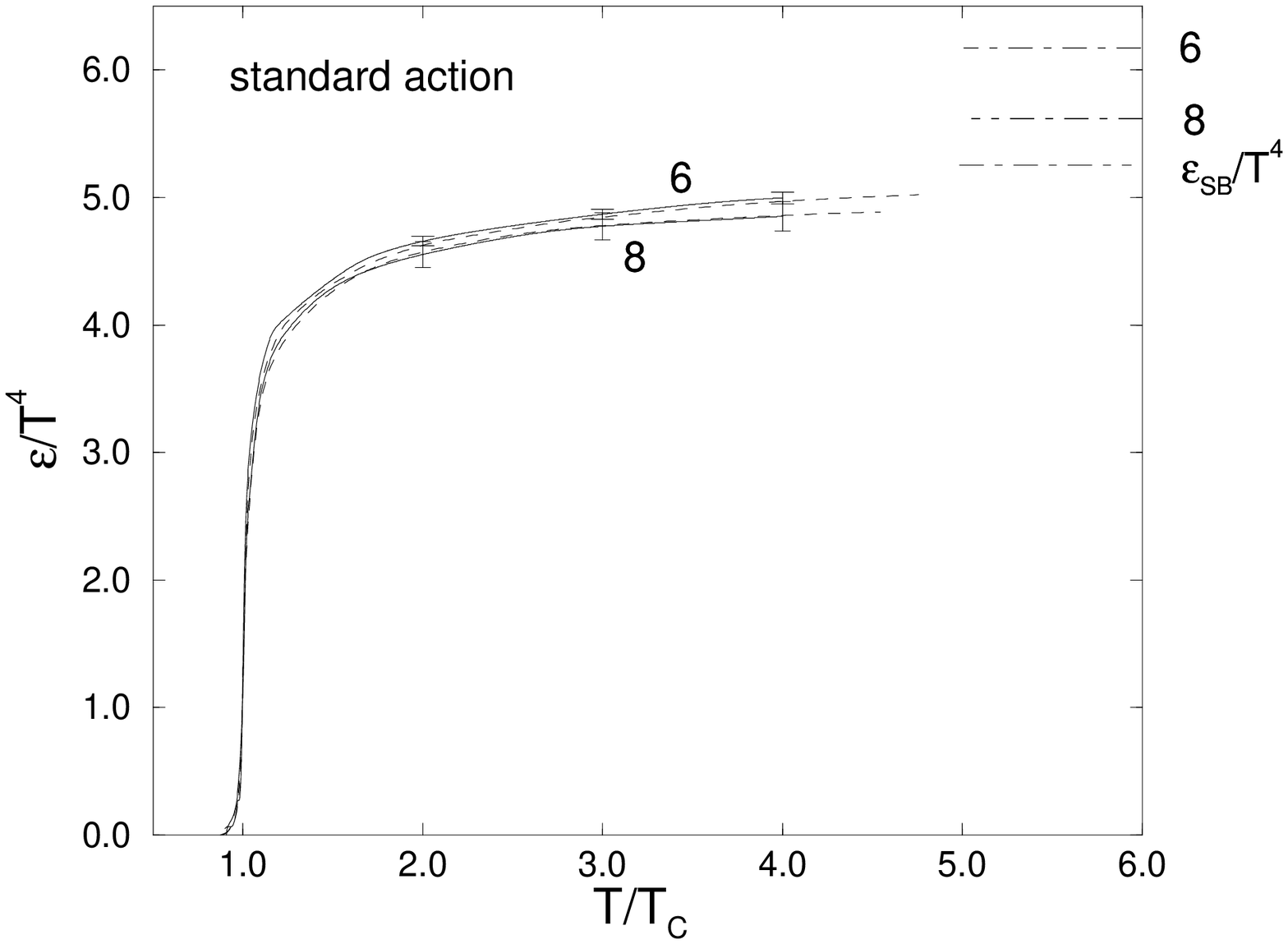}
\end{center}
\caption{Energy density (lower figure) 
and pressure (upper figure) for the plaquette action 
for $N_t=6$ and 8.
Solid lines uses the Allton's parametrization of $\sigma$ 
for scale and dashed lines are original results of the Bielefeld group 
using a different scale fixing scheme. }
\label{fig:stresult}
\end{figure}

We thank S.\ Sakai for communications on the asymmetry 
coefficients for the RG-improved action, F.\ Karsch for allowing 
us to reproduce their results in Fig.~\ref{fig:stresult},
and H.P.\ Shanahan for useful comments.
Valuable discussions with B.\ Petersson and F.\ Karsch are gratefully 
acknowledged. 
Numerical simulations for the present work have been carried out with 
the CP-PACS facility 
at the Center for Computational Physics of the University of Tsukuba.
This work is supported in part by Grants-in-Aid of the Ministry of 
Education (Nos.~6768, 6769, 7034, 09304029, 10640246, 10640248). 
AAK and TM are supported by the 
JSPS Research for Future Program. MO, SE and KN are JSPS Research Fellows.


\newpage

\begin{table} [htb]
\vspace{0mm}
\vspace{5mm}
\begin{center}
  \begin{tabular}{ccr}
    lattice & $\beta$ & \#iterations \\ \hline
      $16^3\times4$  &  2.20 -- 2.30  &  26 000  \\
                     &  2.32 -- 3.20  &  20 000  \\
     $16^4$          &  2.20 -- 3.20  &  10 000  \\
      $32^3\times8$  &  2.60 -- 3.80  &  36 000  \\
     $32^4$          &  2.60 -- 3.80  &  12 000  \\
\end{tabular}
\end{center}
\caption{Statistics of our runs.}
\label{tab:stat}
\end{table}

\begin{table} [tb]
\begin{center}
\begin{tabular}{cllll}
$\beta$ & $16^3 \times 4$ & $16^4$
        & $32^3 \times 8$ & $32^4$           \\
\hline
 2.200 & 11.652966(141) &  11.652657( 87) &                  &                 \\  
 2.250 & 11.856453(192) &  11.855865(135) &                  &                 \\  
 2.270 & 11.933127(336) &  11.931453(117) &                  &                 \\  
 2.300 & 12.053019(249) &  12.039327( 87) &                  &                 \\  
 2.320 & 12.123597(183) &  12.108015(141) &                  &                 \\  
 2.350 & 12.221928(144) &  12.206421( 93) &                  &                 \\  
 2.400 & 12.373140(177) &  12.360168(129) &                  &                 \\  
 2.500 & 12.645009(144) &  12.636627(102) &                  &                 \\  
 2.600 & 12.885558(111) &  12.880320( 93) &  12.8802888(225) & 12.8802498(198) \\  
 2.650 &                &                 &  12.9923472(240) & 12.9923202(252) \\  
 2.700 &                &                 &  13.0990662(657) & 13.0987350(177) \\  
 2.750 &                &                 &  13.2009177(255) & 13.2000711(171) \\  
 2.775 &                &                 &  13.2497922(213) & 13.2489255(162) \\  
 2.800 & 13.299015(138) &  13.296786( 99) &                  &                 \\  
 2.850 &                &                 &  13.3897845(186) & 13.3890450(186) \\  
 2.900 &                &                 &  13.4780103(183) & 13.4773656(201) \\  
 3.000 & 13.644306(141) &  13.643502( 63) &  13.6438083(141) & 13.6433112(162) \\  
 3.200 & 13.938774( 87) &  13.938480( 69) &  13.9385442(174) & 13.9383060(150) \\  
 3.400 &                &                 &  14.1934539(120) & 14.1933099(150) \\  
 3.600 &                &                 &  14.4165000(162) & 14.4164202(114) \\  
 3.800 &                &                 &  14.6135793(141) & 14.6135304(117) \\  
\end{tabular}
\end{center}
\caption{Expectation value of action density $\langle S \rangle$ for 
our runs.}
\label{tab:plaq16}
\end{table}

\begin{table}[tb] \centering
\begin{tabular}{  l  l  c  c  c  }
~$\beta$  & ~~$a\sqrt{\sigma}$ & lattice & \# of conf. & Ref.  \\ \hline
2.1508 & 0.5054(93) & $9^3\times 18 $ & 400 &\cite{ourPot97}\\
2.2827 & 0.3864(32) & $12^3\times 24$ & 200 &\cite{ourPot97}\\
2.40   &  0.3096(54) & $16^3\times 32$ & 50 &\cite{kaneko}\\
2.5157 & 0.2559(23) & $18^3\times 36$  & 100 &\cite{ourPot97}\\
2.60    & 0.2313(58)  & $16^3\times 32$ & 50 &\cite{kaneko}\\
2.70    & 0.1963(34)  & $16^3\times 32$ & 100 &\cite{kaneko}\\
3.20    & 0.1029(19) & $32^4$         & 50  &\cite{kaneko}\\
\end{tabular}      
\vskip 2mm
\caption{Results for string tension obtained with the RG improved action.}
\label{tab:allsig}
\vskip 5mm
\end{table}

\begin{table} [htb]
\begin{center}
\begin{tabular}{clll}
$N_s^3\times N_t$ &
 $\beta_c(N_t,N_s)$ & $\beta_c(N_t,\infty)$ & $T_c/\sqrt{\sigma}$
\\
\hline
 $12^3\times 3$ &     2.1528(9)         &  2.1551(12)   &  0.665(10)        \\
 $16^3\times 4$ &     2.2863(10)        &  2.2879(11)   &  0.654(4)         \\
 $18^3\times 6$ &     2.5157(7)         &  2.5206(24)   &  0.654(5)          \\
 $32^3\times 8$ &     2.7103(32)        &  2.7124(34)   &  0.652(6)          \\
\end{tabular}
\end{center}
\caption{Critical coupling of the  deconfinement transition
for the RG improved action on an $N_s^3\times N_t$ lattice, infinite spatial 
volume extrapolation and the ratio $T_c / \protect\sqrt{\sigma} $ 
for infinite spatial volume.  
Allton's parametrization of string tension is employed to fix
the temperature scale. }
\label{tab:critical}
\end{table}

\begin{table}[htb]
\begin{center}
\begin{tabular}{cccccc}
   \multicolumn{6}{ c }{$p(N_t,N_s) / p_{\rm SB} $}\\
\hline
$N_t$  & 4 & 6 & 8 & 10 & 12  \\
\hline
RG   &   0.1971              &     0.7086    &0.8213& 0.8734  &0.9024 \\
plaquette\cite{bielefeld22}  &   1.4833   &1.1697   &1.0748 & 1.0398  &1.0229 \\
\end{tabular}
\end{center}
\caption{Perturbative high temperature limit of pressure on the lattice 
in units of its continuum value ($p_{\rm SB}/T^4 = 8\pi^2/45$).
Results for the case $N_s / N_t = 4$ are listed.}
\label{tab:PHL}
\end{table}

\newpage

\newpage

\newpage

\newpage

\newpage

\newpage

\end{document}